\documentclass[%
amsmath,amssymb,
aps,
pra,superscriptaddress,twocolumn
]{revtex4-2}

\usepackage{amssymb}
\usepackage{graphicx}% Include figure files
\usepackage{dcolumn}% Align table columns on decimal point
\usepackage{bm}% bold math
\usepackage{amsmath}
\usepackage{lipsum}
\usepackage{amsthm}
\usepackage{mathrsfs}
\usepackage{mathbbol}
\usepackage{braket}
\usepackage{physics}
\usepackage{eqparbox}
\usepackage[colorlinks, citecolor=blue, anchorcolor=blue, linkcolor=blue,urlcolor=blue]{hyperref}% add hypertext capabilities

\begin{document}

\title{Sound propagation in one-dimensional quantum droplets}

\affiliation{Zhejiang Key Laboratory of Quantum State Control and Optical Field Manipulation, Department of Physics, Zhejiang Sci-Tech University, Hangzhou 310018, China}
\affiliation{Lanzhou Center for Theoretical Physics, Key Laboratory of Theoretical Physics of Gansu Province, Key Laboratory of Quantum Theory and Applications of MoE, Gansu Provincial Research Center for Basic Disciplines of Quantum Physics, Lanzhou University, Lanzhou 730000, China}

\author{Zizhou Yuan}
\affiliation{Zhejiang Key Laboratory of Quantum State Control and Optical Field Manipulation, Department of Physics, Zhejiang Sci-Tech University, Hangzhou 310018, China}

\author{Jiarui Xiao}
\affiliation{Zhejiang Key Laboratory of Quantum State Control and Optical Field Manipulation, Department of Physics, Zhejiang Sci-Tech University, Hangzhou 310018, China}

\author{Xiao-Long Chen}\email{xiaolongchen@zstu.edu.cn}
\affiliation{Zhejiang Key Laboratory of Quantum State Control and Optical Field Manipulation, Department of Physics, Zhejiang Sci-Tech University, Hangzhou 310018, China}
\affiliation{Lanzhou Center for Theoretical Physics, Key Laboratory of Theoretical Physics of Gansu Province, Key Laboratory of Quantum Theory and Applications of MoE, Gansu Provincial Research Center for Basic Disciplines of Quantum Physics, Lanzhou University, Lanzhou 730000, China}

\date{\today}

%%%%%%%%%%%%%%%%%%%%%%%%%%%%%%%%%%%%%%%%%%%%%%%%%%
%%%%%%%%%%%%%%%%%%%%%%%%%%%%%%%%%%%%%%%%%%%%%%%%%%
\begin{abstract}
Sound propagation in quantum droplets differs from that in conventional Bose-Einstein condensates (BECs) because of their self-bound nature and the role of quantum fluctuations. We investigate sound propagation in one-dimensional quantum droplets formed by a symmetric Bose-Bose mixture, focusing on finite-size and confinement effects. Using the extended Gross-Pitaevskii equation, we extract the sound velocity from the real-time propagation of localized density perturbations and compare it with the low-energy excitation spectrum. We find that, unlike in a conventional BEC, the sound velocity of a finite droplet is strongly affected by its density profile and quantum-pressure contribution. It decreases with increasing particle number as the droplet evolves from a Gaussian-like to a flat-top profile, approaching the bulk quantum-droplet value. In contrast, external harmonic confinement compresses the droplet and enhances the sound velocity, driving the system toward the acoustic behavior of a trapped BEC. Our results establish sound propagation as a sensitive probe of finite-size effects and the crossover between self-bound quantum droplets and conventional Bose gases, and suggest a feasible route for experimental observation in ultracold $^{39}$K droplets.
\end{abstract}

\maketitle
%\tableofcontents

% I. %%%%%%%%%%%%%%%%%%%%%%%%%%%%%%%%%%%%%%%%%%%%%
%%%%%%%%%%%%%%%%%%%%%%%%%%%%%%%%%%%%%%%%%%%%%%%%%%
\section{INTRODUCTION}
\label{sec:introduction}

Quantum droplets represent a novel class of self-bound quantum fluids stabilized by quantum fluctuations beyond the mean-field description. Unlike conventional dilute Bose-Einstein condensates (BECs) which usually utilize external confinement to maintain stability, quantum droplets can sustain a finite equilibrium density in free space due to the competition between mean-field interactions and beyond-mean-field corrections. In 2015, Petrov demonstrated that the repulsive Lee-Huang-Yang (LHY) correction arising from quantum fluctuations can stabilize a self-bound ultradilute phase for binary Bose mixtures~\cite{petrov2015quantum,petrov2016ultradilute}. 
The first experimental realization of quantum droplets was achieved in dipolar gases, where the interplay between anisotropic dipole-dipole interactions and quantum fluctuations leads to a self-bound state in strongly magnetic atoms, such as $^{164}$Dy and $^{166}$Er~\cite{kadau2016observing,ferrier2016observation,schmitt2016self,chomaz2016quantum}. This discovery opened a new avenue for studying quantum fluids with competing interactions and stimulated extensive investigations of collective excitations, stability, and supersolid behavior in dipolar systems~\cite{tanzi2019supersolid,bottcher2019dilute,tanzi2019observation}.
Shortly afterwards, quantum droplets were also realized in Bose-Bose mixtures, which were first predicted to stabilize an ultradilute self-bound phase~\cite{petrov2015quantum}. Experimental observations in homonuclear and heteronuclear mixtures, including $^{39}$K mixtures~\cite{cabrera2018quantum,semeghini2018self,cheiney2018bright,ferioli2019collisions,skov2021observation}, $^{41}$K-$^{87}$Rb~\cite{derrico2019observation,cavicchioli2025dynamical}, and $^{23}$Na-$^{87}$Rb~\cite{guo2021lhy}, have provided a highly tunable platform for investigating quantum fluctuation effects in quantum gases~\cite{astrakharchik2018dynamics,tylutki2020collective,hu2020collective,guo2021lhy,sturmer2021breathing,cikojevic2021dynamics,hu2022collisional,dong2022internal,du2023ground,fei2024collective,charalampidis2025twocomponent,xiao2026one}.
More recently, the observation of droplet states in ultracold polar molecular gases, realized in $\mathrm{NaCs}$~\cite{bigagli2024observation} and $\mathrm{NaRb}$~\cite{shi2026bose}, has extended quantum droplet physics toward strongly interacting molecular quantum systems~\cite{zhang2026observation}.

Collective excitations and sound propagation are among the most direct probes of the equation of state and low-energy dynamics of quantum fluids. In ultracold atomic gases, sound velocities have been extensively measured by creating localized density perturbations and monitoring their propagation dynamics. 
Andrews \textit{et al.} first observed sound propagation in an elongated BEC by creating a local density perturbation with a focused laser beam and monitoring the subsequent wave-packet evolution using nondestructive imaging~\cite{andrews1997propagation}. Similar techniques have subsequently enabled measurements of sound velocities in Fermi gases~\cite{joseph2007measurement}, Bose-Fermi mixtures~\cite{patel2023sound}, and binary Bose superfluids, revealing distinct density and spin sound modes~\cite{kim2020observation}. Theoretical studies have also clarified the dependence of sound velocities on dimensionality, confinement, interaction strength, and temperature, including the observation and interpretation of first and second sound modes in quantum gases~\cite{zaremba1998sound,kavoulakis1998quasi,capuzzi2006sound,meppelink2009sound,ville2018sound,sidorenkov2013second,hoffmann2021second}. More recently, sound excitations have also been investigated in homogeneous Bose gases and supersolid-like quantum gases, highlighting sound propagation as an essential tool for characterizing emergent quantum phases~\cite{christodoulou2021observation,hilker2022first,sindik2024sound,poli2026sound}.

Despite these advances, the understanding of sound propagation and direct sound-velocity measurements in quantum droplets remains incomplete. Previous studies of droplet sound modes have mainly focused on the stability of the phonon excitation spectrum predicted by conventional Bogoliubov theory in the droplet regime. In particular, the long-wavelength density excitation of a binary Bose mixture can acquire an imaginary sound velocity within the mean-field description, while quantum fluctuation effects restore a stable phonon branch and determine the sound velocity of the self-bound state~\cite{petrov2015quantum,ota2020beyond,gu2020phonon,xiong2022effective,rakhimov2026resolving}. Related investigations have been extended to low-dimensional Bose-Bose droplets and dipolar droplets with anisotropic sound properties~\cite{petrov2016ultradilute,boudjemaa2021many,zhang2022phonon}. However, most existing studies have characterized sound velocity from the excitation spectrum or thermodynamic relations, whereas direct dynamical excitation and measurement of sound propagation in finite quantum droplets have not yet been explored. In particular, the effects of finite particle number, density inhomogeneity, and external confinement on the sound velocity extracted from propagation dynamics remain largely unexplored.

In this work, we investigate sound propagation in one-dimensional quantum droplets formed by a symmetric Bose-Bose mixture using time-dependent extended Gross-Pitaevskii simulations, linear excitation analysis, and a sum-rule approach. By creating a localized density perturbation and monitoring its subsequent evolution, we directly extract the sound velocity from the propagation dynamics. 
We show that the sound velocity exhibits a strong dependence on droplet size and gradually approaches the bulk value in the large-particle-number limit. The influence of external harmonic confinement is also investigated, revealing the crossover between self-bound droplet behavior and trapped Bose gases. Furthermore, the dynamical sound velocities are compared with the low-energy excitation spectrum and sum-rule predictions, establishing the connection between real-time propagation and equilibrium collective properties. Our results provide a quantitative characterization of sound propagation in quantum droplets and suggest feasible experimental schemes for observing acoustic excitations in quasi-one-dimensional ultracold $^{39}$K gases.

The remainder of this paper is organized as follows. Section~\ref{sec:theory} introduces the theoretical model, including the dimensionless extended GPE description, the linearization technique, and the sum-rule approach. Section~\ref{sec:results} presents the sound-propagation dynamics, the dependence of sound velocity on particle number and external confinement, and the comparison with collective excitation spectra. The experimental relevance and possible observation schemes are also discussed. Finally, Section~\ref{sec:conclusion} summarizes our main results and outlook.

% II. %%%%%%%%%%%%%%%%%%%%%%%%%%%%%%%%%%%%%%%%%%%%
%%%%%%%%%%%%%%%%%%%%%%%%%%%%%%%%%%%%%%%%%%%%%%%%%%
\section{THEORETICAL FRAMEWORK}
\label{sec:theory}

% 2.1 ———————————————————
\subsection{Effective one-dimensional model and sound excitation}
\label{subsec:model}

We consider a self-bound quantum droplet formed in a one-dimensional (1D) symmetrically weakly interacting Bose-Bose mixture. The two components are assumed to have equal atomic masses, $m_1=m_2=m$, and this 1D droplet originates from the competition between repulsive mean-field interactions and attractive beyond-mean-field quantum fluctuations~\cite{petrov2016ultradilute}. The effective 1D energy density of the binary mixture is given by~\cite{petrov2016ultradilute}
\begin{eqnarray} \label{eq:energyfunctional}
    E^\mathrm{(1D)}/V &=& \frac{(n_1 \sqrt{g_1} -n_2 \sqrt{g_2})^2}{2} 
    + \frac{g\delta g(n_1 \sqrt{g_2} +n_2 \sqrt{g_1})^2}{(g_{1}+g_2)^2} \nonumber 
    \\
    &&- \frac{2\sqrt{m}(n_1 g_1 +n_2 g_2)^{3/2}}{3\pi \hbar},
\end{eqnarray}
where $\delta g=g+g_{12}>0$ characterizes the deviation from the mean-field collapse threshold with $g=\sqrt{g_1g_2}$. The first two energy terms describe the mean-field contribution, while the last term represents the LHY correction arising from quantum fluctuations.

For weakly interacting droplets near the collapse boundary, the mean-field energy imposes the density-locking condition $n_1/n_2=\sqrt{g_2/g_1}$, which allows the binary mixture to be mapped onto an effective single-component description. Introducing the macroscopic wave function $\psi(x)$ via $\psi_\sigma = \left( \frac{\sqrt{g_{\bar{\sigma}}}}{\sqrt{g_1}+\sqrt{g_2}} \right)^{1/2} \psi$ with the spin index $\sigma\neq\bar{\sigma} \in \{1,2\}$ and $|\psi|^2=|\psi_1|^2+|\psi_2|^2$, the total energy functional becomes
\begin{eqnarray} \label{eq:energy_functional}
    E[\psi] &=& \int \mathrm{d}x \bigg[ 
    \frac{\hbar^2}{2m} \left|\frac{\partial \psi}{\partial x}\right|^2 
    + V_{\mathrm{ext}}(x)|\psi|^2 + \frac{g \delta g}{4\mathcal{G}}|\psi|^4  \nonumber \\
    && - \frac{2\sqrt{m}}{3\pi\hbar} g^{3/2}|\psi|^3 
    \bigg],
\end{eqnarray}
where $\mathcal G=\frac{(\sqrt{g_1}+\sqrt{g_2})^2}{4}$ and external harmonic trapping potential $V_{\mathrm{ext}}(x) = \frac{1}{2}m\omega_x^2 x^2$.
The corresponding time-dependent extended Gross-Pitaevskii equation is obtained from the variational principle to Eq.~\eqref{eq:energy_functional}, and takes the form
\begin{equation}  \label{eq:egpe_dimensional}
    i\hbar \frac{\partial \psi}{\partial t} = 
    \bigg[ -\frac{\hbar^2}{2m} \frac{\partial^2}{\partial x^2} 
    + V_{\mathrm{ext}}(x) 
    + \frac{g \delta g}{2\mathcal{G}}|\psi|^2 
    - \frac{\sqrt{m}}{\pi\hbar}g^{3/2}|\psi| 
    \bigg] \psi.
\end{equation}

To eliminate system-dependent parameters and emphasize the intrinsic properties of the droplet, we introduce characteristic units of wave function, length, energy, particle number, velocity~\cite{astrakharchik2018dynamics,tylutki2020collective,du2023ground}, i.e., $\psi_0 = \frac{2\mathcal{G}\sqrt{mg}}{\pi\hbar\delta g}$, $x_0 = \frac{\pi\hbar^2}{mg}\sqrt{\frac{\delta g}{2\mathcal{G}}}$, $E_0=\hbar\omega_0=\hbar^2/(m x_0^2)$, $N_0=\psi_0^2 x_0$, $v_0 = \frac{\hbar}{m x_0}=\frac{g}{\pi\hbar}\sqrt{\frac{2\mathcal{G}}{\delta g}}$.
After the rescaling procedure described above, the dynamics are governed by the dimensionless equation
\begin{equation} \label{eq:egpe_dimensionless}
    i \frac{\partial \psi}{\partial t} = 
    \left[ -\frac{1}{2} \frac{\partial^2}{\partial x^2} 
    + \frac{1}{2}\kappa^2 x^2
    + |\psi|^2 - |\psi| \right] \psi,
\end{equation}
where dimensionless $\kappa=\omega_x/\omega_0$ denotes the confinement strength and the normalization condition becomes $\int |\psi(x)|^2dx=N/N_0$. Equation~\eqref{eq:egpe_dimensionless} provides the basis for both equilibrium and dynamical simulations. The ground state is obtained by imaginary-time evolution, while sound propagation is investigated through real-time evolution following a controlled density excitation.

To generate a localized acoustic excitation, we introduce a weak Gaussian repulsive barrier with the characteristic height $A_{\mathrm{bar}}>0$ and width $\sigma_\mathrm{bar}$
\begin{equation} \label{eq:gaussian_barrier}
V_{\mathrm{bar}}(x)=A_{\mathrm{bar}}\exp\left(-\frac{x^2}{2\sigma_{\mathrm{bar}}^2}\right),
\end{equation}
which creates a density depletion or dip at the center $x=0$ of the droplet. After preparing the equilibrium state $\psi_{\mathrm{dip}}(x)$ in the presence of the barrier, the potential is suddenly removed at $t=0$, and the system evolves via Eq.~\eqref{eq:egpe_dimensionless} within the linear-response regime. The induced density perturbation then splits into two counter-propagating waves. Thus, the dynamical sound velocity is extracted from the trajectory of the density minimum $x_{\mathrm{dip}}(t)$, i.e., 
\begin{equation}
 c(t)=\left|\frac{dx_\mathrm{dip}}{dt}\right|,    
\end{equation}
where the final sound velocity is determined from the region in which the propagation becomes approximately linear and constant.

% 2.2 ———————————————————
\subsection{Collective excitation spectrum and analytic sound velocity}
\label{subsec:linearization}

The sound velocity extracted from the real-time propagation provides a direct measurement of the acoustic response of a finite droplet. To identify its relation with the intrinsic collective excitations, we further analyze the low-energy excitation spectrum by linearizing the dimensionless extended Gross-Pitaevskii equation around the equilibrium state~\cite{tylutki2020collective,hu2020collective,dong2022internal,du2023ground,fei2024collective,charalampidis2025twocomponent}.

Following the standard procedure, we linearize the time-dependent Bose field in the dimensionless Eq.~\eqref{eq:egpe_dimensionless} as a superposition of a static classical field and a small-amplitude fluctuation
\begin{equation}
\psi(x,t) = e^{-i\mu t}\left[ \phi_0(x) + \delta\phi(x,t) \right],
\end{equation}
where $\mu$ is the chemical potential and $\phi_0(x)$ is the real-valued stationary wave function. The fluctuation is further expanded as
\begin{equation}
\delta\phi(x,t) = \sum_j \left[ u_j(x) e^{-i\omega_j t} + v_j^*(x) e^{i\omega_j t} \right],
\end{equation}
where $\{u_j(x),v_j(x)\}$ are the corresponding amplitudes of the $j$-th quasiparticle with the excitation frequency $\omega_j$. After substituting this linearized expression back into the time-dependent extended GPE and retaining terms up to first order in the small fluctuations $\{u_j(x),v_j(x)\}$, we obtain the stationary equation for the ground-state wave function $\phi_0(x)$, as well as the coupled equations for the small fluctuations, i.e.,
\begin{equation} \label{eq:linearization}
\begin{pmatrix}
\mathcal{H} - \mu + \mathcal{M} & \mathcal{M} \\
-\mathcal{M} & -(\mathcal{H} - \mu + \mathcal{M})
\end{pmatrix}
\begin{pmatrix} u_j \\ v_j \end{pmatrix}
= \omega_j
\begin{pmatrix} u_j \\ v_j \end{pmatrix},
\end{equation}
with the elements $\mathcal{H} = -\frac{1}{2}\frac{\partial^2}{\partial x^2} + V_{\mathrm{ext}}(x) + \phi_0^2 - \phi_0$ and $\mathcal{M} = \phi_0^2 - \frac{1}{2}\phi_0$. 
After obtaining the stationary wave function $\phi_0(x)$, Eq.~\eqref{eq:linearization} can be further diagonalized to extract the excitation spectrum $\omega_j$. The lowest compressional excitation corresponds to the long-wavelength density mode and provides an independent estimate of the sound velocity. Moreover, these low‑lying excitations offer a complementary check on the sound-velocity behavior obtained from the real-time dynamical evolution.

For a homogeneous system ($V_\mathrm{ext}=0$), the ground-state wave function is uniform, which simplifies the chemical potential to $\mu = \phi_0^2 - \phi_0$. 
The mode amplitudes can be expanded with the plane waves as $\{u_j,v_j\}(x) = \{u_k,v_k\} e^{ikx}$, and the coupled Eqs.~\eqref{eq:linearization} yield the dispersion relation
\begin{equation} \label{eq:dispersion}
\omega(k) = \sqrt{\epsilon_k^2 + 2\mathcal{M}\epsilon_k},
\end{equation}
with $\epsilon_k=k^2/2$. In the long-wavelength limit, i.e., $k\to 0$, the linear phonon dispersion follows $\omega(k)=\tilde{c}k \approx\sqrt{\mathcal{M}}k$ with the dimensionless sound velocity $\tilde{c}$ being $\tilde{c} = \sqrt{\phi_0^2 - \frac{1}{2}\phi_0}$.
In terms of the ground-state density $n$ and physical units, we obtain the compact analytical expression of the sound velocity in a uniform quantum droplet
\begin{equation} \label{eq:sound_analytical}
c_\mathrm{QD}^\mathrm{analytical} = \sqrt{\frac{g \delta g}{2\mathcal{G}} \frac{n}{m} - \frac{g^{3/2}}{2\pi\hbar}\sqrt{\frac{n}{m}}}.
\end{equation}
Notably, this expression applies to both symmetric ($g_1=g_2$) and asymmetric ($g_1\neq g_2$) binary mixtures~\footnote{For one-dimensional symmetric droplets, we note that our analytical sound velocity [Eq.~\eqref{eq:sound_analytical}] differs in form from the density-channel result derived from thermodynamic compressibility for a two-component mixture [Eq. (20a) of Ref.~\cite{ota2020beyond}]; nevertheless, the two expressions agree to within $1\%$ under the same parameters.
This minor deviation stems from different treatments of the LHY term. We adopt a first-order expansion in the small parameter $\delta g/g$ near the collapse threshold, whereas Ref.~\cite{ota2020beyond} retains the full dependence. The close agreement confirms the reliability of our effective single-mode description in the homogeneous limit.}. In finite droplets, however, the broken translational symmetry gives rise to discrete collective modes rather than a continuous phonon spectrum. We therefore use the lowest density excitation obtained from the linearization analysis as a characteristic velocity scale, and compare it with that measured directly from real-time propagation.

% 2.3 ———————————————————
\subsection{Sum-rule analysis of the collective density mode}
\label{subsec:sum-rule}

As a complement to the real-time propagation and linearization analysis, we employ the sum-rule approach~\cite{stringari1996collective,menotti2002collective,dalfovo1999theory,pitaevskii2016bose} to characterize the low-energy density response of the droplet. This approach relies on the static ground-state properties and provides an analytical upper-bound estimate that can be directly compared with our numerical results.

Previous studies have shown that the ground-state wave function of a quantum-fluctuation-stabilized droplet can deviate significantly from a conventional Gaussian shape, exhibiting a flat-top density distribution at large atom numbers~\cite{astrakharchik2018dynamics,otajonov2019stationary,sturmer2021breathing,du2023ground,fei2024collective,xiao2026one}. Therefore, to accurately capture both the Gaussian and flat-top regimes, we introduce a normalized super-Gaussian variational ansatz~\cite{otajonov2019stationary,sturmer2021breathing,du2023ground,fei2024collective,xiao2026one}
\begin{equation} \label{eq:super_gaussian_ansatz}
\phi_0(x) = \sqrt{\frac{\alpha N}{\sigma\Gamma\left(\frac{1}{2\alpha}\right)}} \exp\left[-\frac{1}{2}\left(\frac{x}{\sigma}\right)^{2\alpha}\right],
\end{equation}
with two variational variables, i.e., the cloud width $\sigma$ and the shape exponent $\alpha$, and the Gamma function $\Gamma(x)$ for normalization. The width sets the spatial extent of the droplet, while the exponent governs the density structure, ranging from Gaussian-like to flat-top. For $\alpha=1$, the ansatz reduces to the standard Gaussian form, whereas $\alpha>1$ can describe an increasingly flat central plateau.

Substituting this ansatz into the energy functional and integrating over space, we obtain the energy per particle $\epsilon_{\mathrm{tot}} = E/N$ as
\begin{eqnarray} \label{eq:energy_super_gaussian}
\epsilon_{\mathrm{tot}}(\sigma,\alpha)
&=&\frac{\hbar^2\alpha^2}{2m\sigma^2}\frac{\Gamma\left(2-\frac{1}{2\alpha}\right)}{\Gamma\left(\frac{1}{2\alpha}\right)}+\frac{m\omega_x^2\sigma^2}{2}\frac{\Gamma\left(\frac{3}{2\alpha}\right)}{\Gamma\left(\frac{1}{2\alpha}\right)} \nonumber
\\
&&+\frac{g\delta g}{4\mathcal{G}}\left(\frac{1}{2}\right)^{\frac{1}{2\alpha}}\frac{\alpha N}{\sigma\Gamma\left(\frac{1}{2\alpha}\right)} \nonumber
\\ 
&&-\frac{2\sqrt{m}g^{3/2}}{3\pi\hbar}\left(\frac{2}{3}\right)^{\frac{1}{2\alpha}}\sqrt{\frac{\alpha N}{\sigma\Gamma\left(\frac{1}{2\alpha}\right)}}.
\end{eqnarray}
Hence, by minimizing the energy with respect to the variational parameters, the extremum point $(\alpha_0,\sigma_0)$ is determined via the stationary conditions $\partial\epsilon_{\mathrm{tot}}/\partial\alpha=0$ and $\partial\epsilon_{\mathrm{tot}}/\partial\sigma=0$, and the wave function can be reconstructed.

In terms of the ground-state wave function within the variational approximation, we turn to estimate the excitation frequency of the breathing mode using the sum-rule approach.
For the breathing excitation, we choose the density operator $F=\sum_i x_i^2$, which couples strongly to the lowest compressional mode. 
The excitation frequency can be estimated by the ratio of the energy-weighted moments $(\hbar\omega_\mathrm{b})^2\leq\frac{m_1(F)}{m_{-1}(F)}$, which can be expressed in terms of equilibrium quantities. The energy-weighted moment $m_{1}(F)=\frac{2\hbar^2}{m}\langle \sum_{i} x_i^2\rangle = \frac{2\hbar^2N\sigma_{0}^2}{m} \frac{\Gamma\left(\frac{3}{2\alpha_0} \right)}{\Gamma\left( \frac{1}{2\alpha_0} \right)}$ can be straightforwardly calculated via the commutator relation as $m_1(F) = \frac{1}{2}\langle \psi | [[F, H], F] | \psi \rangle$.
Moreover, the inverse-energy-weighted moment $m_{-1}$ can be evaluated by the relation with the static polarizability $\chi$ as $m_{-1}(F)=\frac{1}{2}\chi= -\frac{N}{m} \frac{\partial \langle x^2 \rangle}{\partial \omega_x^2}$~\cite{stringari1996collective,menotti2002collective,pitaevskii2016bose}. 
After simple algebra, the breathing-mode frequency can be estimated by $\omega_\mathrm{b}^2 = \frac{m_1}{m_{-1}\hbar^2} = -\sigma_0\frac{\partial \omega_x^2}{\partial \sigma_0}$, which gives the analytic expression
\begin{eqnarray} \label{eq:omegab}
\omega_\mathrm{b}^2 &=&
4\omega_x^2-\frac{g\delta g}{4m\mathcal{G}}\left(\frac{1}{2}\right)^{\frac{1}{2\alpha_0}}\frac{\alpha_0 N}{\sigma_0^3\Gamma\left(\frac{3}{2\alpha_0}\right)}\nonumber
\\
&& +\frac{g^{3/2}}{2\pi\hbar\sqrt{m}}\left(\frac{2}{3}\right)^{\frac{1}{2\alpha_0}}\frac{\sqrt{\alpha_0 N\Gamma\left(\frac{1}{2\alpha_0}\right)}}{\sigma_0^{5/2}\Gamma\left(\frac{3}{2\alpha_0}\right)}.
\end{eqnarray}
In the non-interacting limit ($\alpha_0=1$ or $N=0$), the breathing-mode frequency reduces to the harmonic-oscillator value $\omega_\mathrm{b} =
2\omega_x$ for a standard Gaussian wave function. This analytical estimate provides a useful comparison with the numerical breathing frequencies obtained from the linearization technique. Moreover, although the sum-rule approach does not describe the propagation process explicitly, it offers an equilibrium characterization of the acoustic response and thus provides an additional independent check on the dynamical sound velocity.

% III. %%%%%%%%%%%%%%%%%%%%%%%%%%%%%%%%%%%%%%%%%%%
%%%%%%%%%%%%%%%%%%%%%%%%%%%%%%%%%%%%%%%%%%%%%%%%%%
\section{RESULTS AND DISCUSSIONS} \label{sec:results}

In this section, we investigate the acoustic response of one-dimensional quantum droplets in Bose-Bose mixtures~\cite{cabrera2018quantum,semeghini2018self,cheiney2018bright}, by combining real-time density dynamics with collective excitation analysis. We consider a quasi-one-dimensional symmetric $^{39}$K Bose-Bose mixture with realistic confinement parameters $(\omega_x,\omega_\perp)=2\pi\times[300,1.85\times10^4]$ Hz~\cite{greiner2001exploring}, where the strong transverse confinement ensures the validity of the one-dimensional description~\cite{gorlitz2001realization,schreck2001quasipure,greiner2001exploring}. 
Under tight transverse confinement ($\omega_{\perp}\gg\omega_x$), the effective 1D interaction strength takes the form $g^\mathrm{(1D)}\approx\frac{4\hbar^{2}a_s}{ml_{\perp}^{2}}(1-1.4603\frac{a_s}{l_{\perp}})^{-1}$~\cite{olshanii1998atomic}, or $g^\mathrm{(1D)}\approx\frac{\hbar^2a_s}{2ml_{\perp}^2}$~\cite{petrov2016ultradilute}, with the oscillator length $l_{\perp}$ in the tightly confined $y$-$z$ plane. For our simulations, we choose the droplet regime with the dimensionless effective 1D interaction strengths $g_1=g_2=g=0.100$ and $g_{12} = -0.099$, ensuring $\delta g/g \ll 1$. These values correspond to 3D scattering lengths $a_{s,11}=a_{s,22}=56.99a_0$, and $a_{s,12}=-56.42a_0$ via $g^\mathrm{(1D)}\approx\frac{\hbar^2a_s}{2ml_{\perp}^2}$ that can be tuned using the magnetic Feshbach resonance of $^{39}$K atoms.
The scaled particle number is chosen in the range $N/N_0\in[2,20]$, corresponding to realistic atom number $N$ ranging from $1801$ to $18006$, which covers the crossover from compact Gaussian-like droplets to extended flat-top droplets. The dimensionless axial confinement strength is varied within $\kappa\in[0,0.2]$ to examine the transition from self-bound droplets to trapped Bose gases~\cite{astrakharchik2018dynamics,du2023ground}. To distinguish finite-size and confinement effects on the sound velocity, we separately study the dependence on particle number $N$ at fixed $\kappa$ and on confinement strength $\kappa$ at fixed $N$.

% 3.1 ——————————————————— GS
\subsection{Ground-state properties and droplet morphology}
\label{subsec:groundstate}
\begin{figure}[t]
\centering
\includegraphics[width=0.48\textwidth]{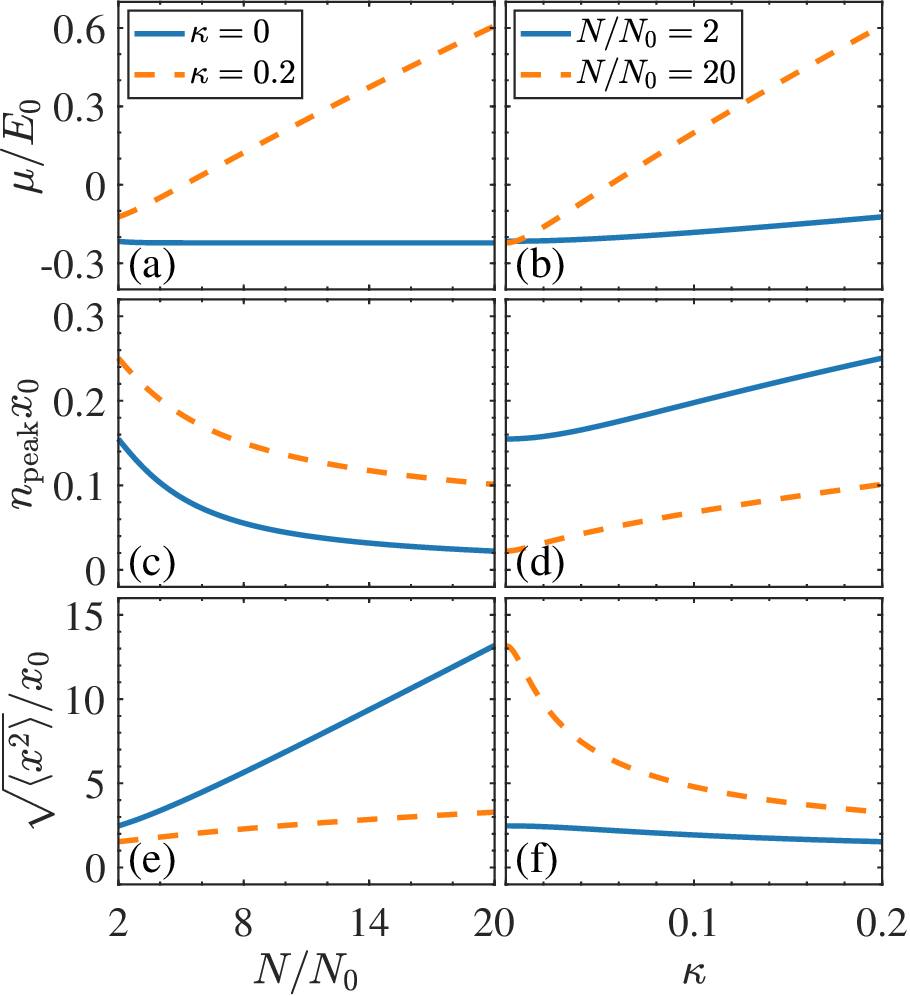}
\caption{Ground-state properties for various values of confinement strength $\kappa$ (left column) and particle number $N/N_0$ (right column). The chemical potential $\mu$, peak density $n_{\mathrm{peak}}$ with total density normalized to unity, and root-mean-square radius $\sqrt{\langle x^2\rangle}$ are plotted from top to bottom. Here, the data for $\kappa=0$ and $N/N_0=2$ are denoted by the blue solid lines, while those for $\kappa=0.2$ and $N/N_0=20$ are shown in orange dashed lines.}
\label{fig1:groundstate}
\end{figure}

Before analyzing the sound propagation dynamics, we first characterize the ground-state properties of the droplets. Figure~\ref{fig1:groundstate} summarizes the chemical potential, normalized peak density, and root-mean-square (RMS) radius as functions of particle number $N/N_0$ and confinement strength $\kappa$. These quantities determine the density profile and spatial scale of the droplet, i.e., compact Gaussian-like and extended flat-top structures, providing the equilibrium background for the subsequent analysis of sound propagation.

As shown in Fig.~\ref{fig1:groundstate}(a) and Fig.~\ref{fig1:groundstate}(b), the chemical potential $\mu$ exhibits distinct behaviors between free-space and trapped droplets. In the absence of external confinement, $\mu$ remains negative and gradually approaches the uniform droplet limit $\mu=-\frac{2}{9}E_0$ with increasing $N/N_0$, as shown in the blue solid line~\cite{petrov2016ultradilute}, reflecting the self-bound nature of the system. 
When the harmonic trap is introduced, the chemical potential is shifted upward due to the additional confinement energy and increases with both particle number $N/N_0$ and confinement strength $\kappa$. These modifications of the energy scale and density profile are expected to influence the density response and sound propagation.

The evolution of the density distribution is further illustrated by the normalized peak density $n_\mathrm{peak}$ in Fig.~\ref{fig1:groundstate}(c) and Fig.~\ref{fig1:groundstate}(d). At fixed confinement strength $\kappa$ shown in the left subplot, the peak density decreases with increasing $N/N_0$, indicating that the droplet expands from a compact Gaussian-like profile toward an extended flat-top structure. In contrast, increasing $\kappa$ compresses the droplet and enhances the central density, as shown in the right subplot. 

The corresponding changes in droplet size are reflected in the RMS radius shown in Fig.~\ref{fig1:groundstate}(e) and Fig.~\ref{fig1:groundstate}(f). In free space, the radius denoted by the blue solid line increases rapidly with particle number due to the formation of a broader flat-top region, whereas the axial trap suppresses this expansion as shown by the orange dashed line. 
This suppression effect is particularly pronounced for the larger droplet ($N/N_0=20$), whose extended spatial profile is more sensitive to external confinement, as shown by the orange dashed line in Fig.~\ref{fig1:groundstate}(f).
These results demonstrate that particle number mainly determines the intrinsic size of the droplet, while external confinement modifies the density distribution through compression. Thus, the interplay of these two effects is expected to significantly impact the sound-propagation dynamics discussed below.
\begin{figure*}[t]
\centering
\includegraphics[width=0.94\textwidth]{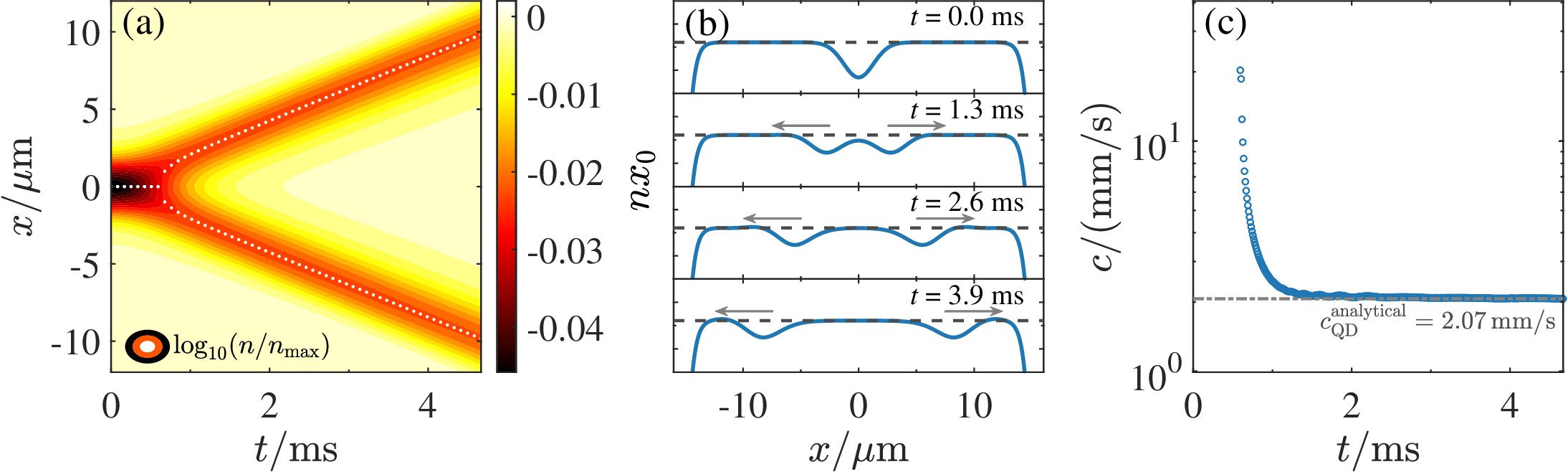}
\caption{Schematic illustration of sound-wave excitation and velocity extraction for $\kappa=0$ and $N/N_0=50$. 
(a) Spatiotemporal contour plot of the total density $n(x,t)$ with the color gradient scaled by $\log_{10}(n/n_{\max})$. The density evolution shows the splitting of the initial density dip into two outward-propagating dips indicated by white dots.
(b) Instantaneous density profiles with the dip splitting and moving at various times $t=0$, $1.3$, $2.6$, and $3.9$ ms. The gray dashed lines indicate the unperturbed flat-top density baseline, and gray arrows mark the propagation directions of the density dips. 
(c) Instantaneous sound velocity $c$ obtained from the motion of the density-dip minimum. The gray dash-dotted line denotes the analytical sound velocity of the quantum droplet $c_\mathrm{QD}^\mathrm{analytical}=2.07$ mm/s estimated from Eq.~\eqref{eq:sound_analytical}.}
\label{fig2:excitation}
\end{figure*}

% 3.2 ——————————————————— Sound
\subsection{Sound propagation and velocity extraction}
\label{subsec:sound}

We next investigate real-time sound propagation by introducing a localized density perturbation into the droplet. A repulsive Gaussian barrier is first applied at the droplet center to create a density depletion or dip, and then suddenly removed at $t=0$, as done in cold atomic experiments~\cite{andrews1997propagation,meppelink2009sound,patel2023sound}.
The resulting density perturbation evolves into two counter-propagating density waves, whose motion provides a direct measurement of the sound velocity. 
In practice, the barrier height is set to produce an approximately $10\%$ density dip at the droplet center, reaching a compromise between reliable tracking and avoiding nonlinear effects. The barrier width is chosen to balance dispersion of the density dip and propagation distance, though the extracted sound velocity in the stable regime is largely insensitive to this choice for a fixed dip depth.

In Fig.~\ref{fig2:excitation}, a representative example of sound excitation and propagation dynamics is presented for a free-space droplet with $\kappa=0$ and $N/N_0=50$. In Fig.~\ref{fig2:excitation}(a), a contour plot of the spatiotemporal density evolution shows that the initial density dip rapidly splits into two symmetric density dips propagating outward. The trajectories of the density minima, marked by white dots, become nearly linear after initial splitting, indicating a well-defined propagation velocity. Here, the gradient color scale represents $\log_{10}(n/n_{\max})$, the logarithm of the instantaneous density relative to the unperturbed peak density.
Figure~\ref{fig2:excitation}(b) displays instantaneous density profiles at four selected times from $0$ to $3.9$ ms, showing that the density dips preserve their shape during propagation with only weak dispersion.

Figure~\ref{fig2:excitation}(c) shows the instantaneous sound velocity extracted from the motion of the density-dip minimum. The initial rapid variation corresponds to the transient splitting process and is excluded from the velocity determination. After this stage, the propagation velocity stabilizes and approaches the analytical quantum-droplet sound velocity $c_\mathrm{QD}^\mathrm{analytical}=2.07$ mm/s derived in Eq.~\eqref{eq:sound_analytical} from the long-wavelength excitation spectrum, indicated by the horizontal gray dash-dotted line. 
This excellent agreement confirms that the density-dip propagation provides a reliable dynamical measurement of the sound velocity in finite droplets.

% 3.3 ——————————————————— effects
\subsection{Finite-size and confinement effects on sound velocity}
\label{subsec:effects}

\begin{figure}[t]
\centering
\includegraphics[width=0.48\textwidth]{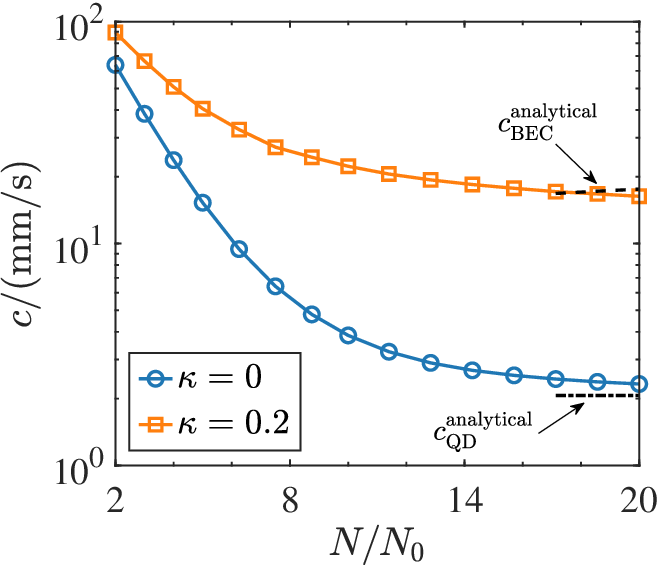}
\caption{Sound velocity $c$ as a function of particle number $N$ for confinement strengths $\kappa=0$ (blue circles) and $\kappa=0.2$ (orange squares). 
The solid lines with hollow symbols represent the results from numerical simulations of the extended GPE, while the black dash-dotted and dashed lines denote the analytical sound velocity for the quantum droplet $c_\mathrm{QD}^\mathrm{analytical}$ in Eq.~\eqref{eq:sound_analytical} for $\kappa=0$ and that for a weakly interacting BEC $c_\mathrm{BEC}^\mathrm{analytical}=\sqrt{g n_c/m}$~\cite{pitaevskii2016bose} for $\kappa=0.2$, respectively.}
\label{fig3:soundvelocity_N} 
\end{figure}

\begin{figure}[t]
\centering
\includegraphics[width=0.48\textwidth]{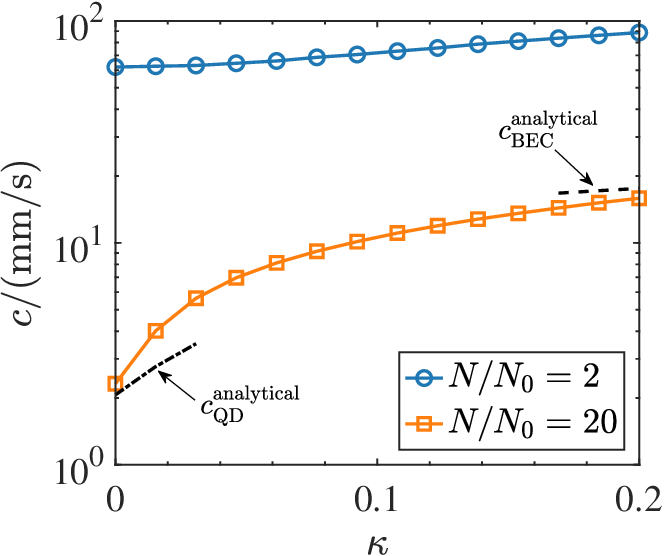}
\caption{Sound velocity $c$ as a function of confinement strength $\kappa$ for particle numbers $N/N_0=2$ (blue circles) and $N/N_0=20$ (orange squares). Here, the black dash-dotted and dashed lines denote the analytical sound velocities of the quantum droplet and the weakly interacting BEC for $N/N_0=20$, respectively. The other properties of the figure are the same as Fig.~\ref{fig3:soundvelocity_N}.}
\label{fig4:soundvelocity_kappa} 
\end{figure}

Having established the dynamical extraction method, we now analyze the dependence of the sound velocity on particle number $N/N_0$ and external confinement strength $\kappa$.
To benchmark the numerical results shown in Fig.~\ref{fig3:soundvelocity_N} and Fig.~\ref{fig4:soundvelocity_kappa}, we include two analytical limit velocities. The analytical quantum-droplet sound velocity $c_\mathrm{QD}^\mathrm{analytical}$ in Eq.~\eqref{eq:sound_analytical} is derived from the homogeneous long-wavelength limit and applies in the large-droplet or weak-trap regimes, while deviations due to finite-size effects are expected for small $N/N_0$. The standard sound velocity $c_\mathrm{BEC}^\mathrm{analytical}=\sqrt{g n_c/m}$~\cite{pitaevskii2016bose} of a conventional weakly interacting single-component $^{39}$K Bose gas is also shown for reference. Here $n_c$ is set equal to the central droplet density, and $g=\frac{4\pi\hbar^2a_s}{m}$ is the 3D interaction strength with the scattering length $a_s=56.99a_0$ equal to the intra-spin scattering length taken for the droplets.

Figure~\ref{fig3:soundvelocity_N} shows the sound velocity $c$ as a function of particle number $N/N_0$ for free-space droplets ($\kappa=0$) and trapped droplets ($\kappa=0.2$). For both cases, the sound velocity decreases monotonically with increasing $N/N_0$. This behavior originates from the morphological evolution of the droplet. At small particle numbers, the droplet possesses a compact Gaussian-like density distribution with strong density gradients. The associated quantum pressure contribution enhances the restoring force against density perturbations~\cite{astrakharchik2018dynamics}, leading to a relatively large sound velocity of about a hundred mm/s. 
With increasing $N/N_0$, the droplet gradually develops an extended flat-top density profile. The central density becomes much smoother, reducing the contribution of quantum pressure, and the sound velocity decreases toward the asymptotic value. 
In the large-particle-number limit, the free-space droplets exhibit a sizable nearly uniform central region, and the numerical results in the blue line with circles converge to the analytical sound velocity $c_\mathrm{QD}^\mathrm{analytical}$ of a homogeneous quantum droplet, indicated by the black dash-dotted line. For finite confinement ($\kappa=0.2$), the significant external trap compresses the density profile and increases the characteristic energy scale. As a result, the sound velocity in the orange line with squares remains relatively higher and approaches the value $c_\mathrm{BEC}^\mathrm{analytical}$ of a weakly interacting single-component Bose gas, denoted by the black dashed line.

In Fig.~\ref{fig4:soundvelocity_kappa}, the dependence of sound velocity $c$ on the external confinement strength $\kappa$ is presented for compact ($N/N_0=2$) and extended ($N/N_0=20$) droplets. The sound velocity increases monotonically with confinement strength in both cases. 
For the compact droplet shown in the blue line with circles, finite-size effects already dominate the acoustic response, so the additional increase induced by the trap is relatively weak. In sharp contrast, the extended flat-top droplet at $N/N_0=20$, indicated by the orange line with squares, is strongly affected by confinement. As the confinement gets stronger (i.e., $\kappa$ rises), the external harmonic potential significantly compresses the broad density distribution, enhances the central density, and thus increases the sound velocity. In the weak-confinement limit, the numerical sound velocity remains close to the quantum-droplet prediction $c_\mathrm{QD}^\mathrm{analytical}$ denoted by the black dash-dotted line.
As confinement becomes relatively stronger, the sound velocity gradually approaches the Bogoliubov sound velocity $c_\mathrm{BEC}^\mathrm{analytical}$ of a weakly interacting Bose gas, indicated by the black dashed line. The behavior of sound velocity indicates a crossover from self-bound droplet behavior toward a trapped BEC-like regime driven by the external confinement.

% 3.4 ——————————————————— Linearization
\subsection{Collective excitation spectrum and sum-rule analysis}
\label{subsec:collective_excitation}

\begin{figure}[t]
\centering
\includegraphics[width=0.48\textwidth]{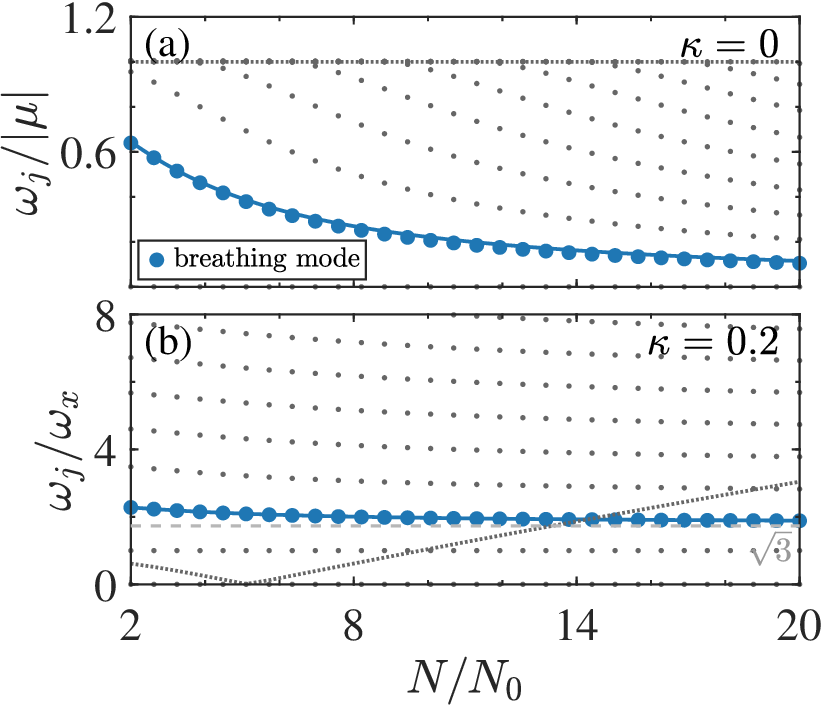}
\caption{Collective excitation frequencies $\omega_j$ as functions of particle number $N/N_0$ for two confinement strengths (a) $\kappa=0$ and (b) $\kappa=0.2$. The excitation frequencies are shown in units of the particle-emission threshold $|\mu|$ and the harmonic trapping frequency $\omega_x$, respectively. 
All dots denote the low-lying excitation spectrum in Eq.~\eqref{eq:linearization} obtained from the linearization technique, and the breathing mode is highlighted by blue circles. For comparison, the breathing-mode frequencies in Eq.~\eqref{eq:omegab} obtained from the sum-rule approach using the super-Gaussian ansatz are denoted by the blue lines.
The gray dotted lines denote the particle-emission threshold $|\mu|$, and the gray dashed line marks the breathing-mode frequency of a 1D BEC $\omega_\mathrm{b}=\sqrt{3}\omega_x$~\cite{stringari1996collective}.}
\label{fig5:spectrum_N}
\end{figure}

\begin{figure}[t]
\centering
\includegraphics[width=0.48\textwidth]{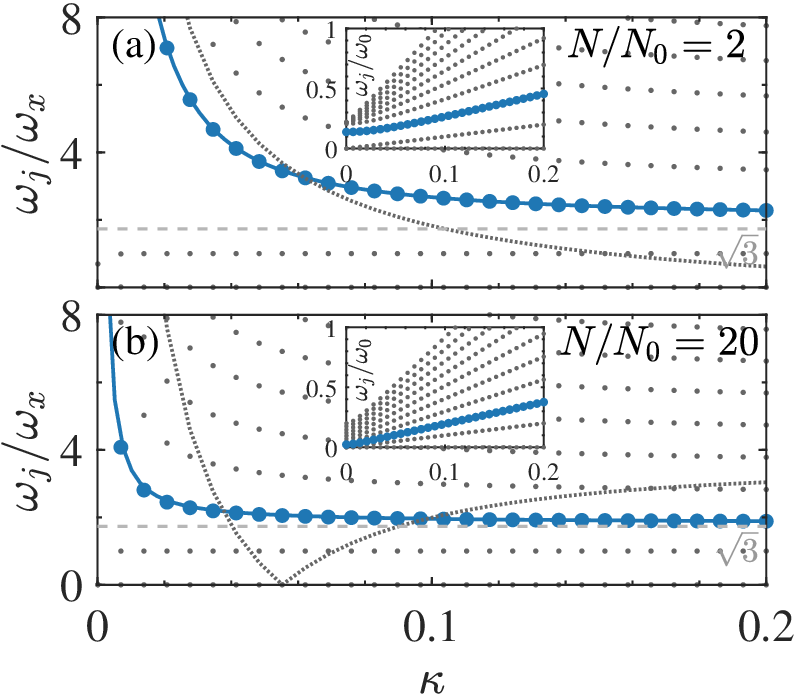}
\caption{Collective excitation frequencies $\omega_j$ as functions of confinement strength $\kappa$ for two particle numbers (a) $N/N_0=2$ and (b) $N/N_0=20$. The frequency unit $\omega_x$ varies with $\kappa$ as $\omega_x=\kappa\omega_0$, and the insets depict the excitation frequencies in the fixed unit $\omega_0$. The
other properties of the figure are the same as Fig.~\ref{fig5:spectrum_N}.}
\label{fig6:spectrum_kappa}
\end{figure}

To further verify the sound-velocity trends from the perspective of microscopic excitations, we calculate the low-energy collective spectrum using the linearization technique in Sec.~\ref{subsec:linearization} and the sum-rule approach in Sec.~\ref{subsec:sum-rule}. The evolution of the collective excitation frequencies with particle number $N/N_0$ and confinement strength $\kappa$ is shown in Fig.~\ref{fig5:spectrum_N} and Fig.~\ref{fig6:spectrum_kappa}, respectively. The excitation spectra obtained from the linearization technique reveal a series of discrete branches evolving smoothly with the system parameters (gray dots). The concerned breathing mode is highlighted by filled blue circles, while the blue solid line represents the sum-rule prediction based on the super-Gaussian ansatz. The gray dotted lines denote the chemical potential modulus $|\mu|$, and the discontinuities indicate the positions at which the chemical potential approaches zero.

In the free-space case shown in Fig.~\ref{fig5:spectrum_N}(a), the excitation frequencies are scaled by the particle-emission threshold $|\mu|$. The low-lying modes remain below this threshold, indicating that they correspond to bound collective excitations of the self-bound droplet. 
With increasing $N/N_0$, finite-size effects are reduced by the formation of an extended flat-top structure, and the breathing-mode frequency decreases with the particle number, consistent with softening of the sound velocity denoted by the blue line in Fig.~\ref{fig3:soundvelocity_N}.
For trapped droplets, the excitation frequencies are expressed in units of the axial trap frequency $\omega_x$ as shown in Fig.~\ref{fig5:spectrum_N}(b) and Fig.~\ref{fig6:spectrum_kappa}. The lowest dipole mode remains fixed at $\omega_\mathrm{d}=\omega_x$, as required by the Kohn theorem. 
The remaining low-energy excitation branches provide qualitative information about the compressional response of the system. 
The narrowing spectral spacing between low-lying modes with increasing $N/N_0$ in Fig.~\ref{fig5:spectrum_N} signals a lower energy cost for long-wavelength excitations, consistent with the decreasing sound velocity shown in Fig.~\ref{fig3:soundvelocity_N}. Conversely, increasing the confinement strength $\kappa$ widens the gaps between the low-lying excitation modes (see insets of Fig.~\ref{fig6:spectrum_kappa}). This raises the characteristic excitation-energy scale of the system and enhances the sound velocity, as shown in Fig.~\ref{fig4:soundvelocity_kappa}.

Specifically, the breathing-mode frequencies $\omega_\mathrm{b}$ obtained from the linearization analysis (filled blue circles) agree well with the sum-rule predictions (blue lines) based on the super-Gaussian variational ansatz. Under relatively strong confinement, the breathing-mode frequency approaches the 1D weakly interacting Bose-gas limit $\omega_\mathrm{b}=\sqrt{3}\omega_x$~\cite{stringari1996collective}. This convergence confirms that external confinement drives the system continuously from the quantum-droplet regime toward conventional weakly interacting BEC behavior. This converging effect is more evident for the extended droplet with $N/N_0=20$ in Fig.~\ref{fig6:spectrum_kappa}(b), consistent with the analysis of sound velocity in Fig.~\ref{fig4:soundvelocity_kappa} that broader droplets are more sensitive to the external trap.

Together, the dynamical sound propagation and collective excitation analysis establish a consistent picture that the acoustic response of finite quantum droplets is controlled by two competing mechanisms. Increasing particle number reduces finite-size quantum-pressure effects and lowers the sound velocity, while external confinement compresses the droplet and enhances the sound velocity.

% 3.5 ——————————————————— Experiment
\subsection{Experimental feasibility}
\label{subsec:experiment}

The above results demonstrate that sound propagation provides a direct probe of the acoustic properties of finite quantum droplets. We now discuss the experimental feasibility of observing such dynamics in current ultracold-atom platforms. 
A natural experimental realization can be based on the two-component $^{39}$K Bose-Bose mixture, where self-bound quantum droplets have already been experimentally demonstrated in both free-space and confined geometries by tuning the interaction strengths to the droplet regime via magnetic Feshbach resonances~\cite{cabrera2018quantum,semeghini2018self}. The quasi-one-dimensional configuration considered here can be achieved by applying a strong transverse confinement together with a much weaker axial confinement, as in the cold atomic experiment~\cite{greiner2001exploring} with trapping frequencies $(\omega_x,\omega_\perp)=2\pi\times(300,1.85\times10^4)$ Hz, ensuring the validity of the effective one-dimensional description.

After preparing the equilibrium droplet, a localized density perturbation can be generated by applying a focused blue-detuned laser beam at the droplet center as in experiments~\cite{andrews1997propagation,meppelink2009sound,patel2023sound}. The laser-induced repulsive potential creates a controllable density depletion, whose amplitude can be adjusted to remain within the linear response regime.
At $t=0$, the laser barrier is suddenly removed, and the density perturbation evolves into two counter-propagating sound waves along the droplet axis, as shown in Fig.~\ref{fig2:excitation}. The subsequent dynamics can be monitored using high-resolution in-situ imaging~\cite{cabrera2018quantum,cheiney2018bright}.
By extracting the trajectories of the density minima from successive density profiles, the sound propagation velocity can be directly determined without requiring reconstruction of the excitation spectrum.

For the representative case shown in Fig.~\ref{fig2:excitation}, corresponding to $N\approx4.5\times10^4$ atoms, the density dips propagate over a distance of approximately $10~\mu$m within about $4$ ms. These spatial and temporal scales are compatible with current $^{39}$K quantum-droplet experiments, i.e., typically $1.5\sim6~\mu$m and $7\sim25$ ms, where the propagation dynamics can be timely recorded before significant atom loss or droplet decay occurs~\cite{cabrera2018quantum, semeghini2018self,cheiney2018bright}. Therefore, the proposed excitation and imaging protocol provides a realistic route toward the direct observation of sound propagation in quasi-one-dimensional quantum droplets.
Beyond demonstrating experimental feasibility, such a measurement would provide a direct test of how finite-size effects and external confinement modify the acoustic response of self-bound quantum fluids. It would also establish a connection between dynamical sound propagation, collective excitations, and the equation of state of quantum droplets.

% IV. %%%%%%%%%%%%%%%%%%%%%%%%%%%%%%%%%%%%%%%%%%%%
%%%%%%%%%%%%%%%%%%%%%%%%%%%%%%%%%%%%%%%%%%%%%%%%%%
\section{CONCLUSIONS AND OUTLOOKS}
\label{sec:conclusion}

In summary, we have investigated sound propagation in finite one-dimensional quantum droplets formed by a symmetric Bose-Bose mixture. By combining real-time dynamics, linearization analysis, and a sum-rule approach, we have established a direct connection between the propagation of localized density perturbations and the static and collective properties of self-bound quantum fluids. In contrast to the conventional determination of sound velocity from the long-wavelength excitation spectrum of a homogeneous system, our approach allows the acoustic response of a finite and inhomogeneous droplet to be characterized directly in real space and real time.

We first demonstrated that a localized density depletion generated by a repulsive potential barrier evolves into two counter-propagating density waves after the barrier is removed. The propagation velocity extracted from the trajectories of the density minima provides a direct measure of the dynamical sound velocity. 
An important feature revealed by our calculations is the pronounced finite-size dependence of the sound velocity. Small droplets exhibit a strong contribution from quantum pressure and possess a spatially varying density profile, leading to a propagation velocity that differs substantially from the bulk value. With increasing particle number, the droplet develops an extended flat-top structure, and the sound velocity gradually approaches its thermodynamic-limit value. This behavior demonstrates that sound propagation provides a sensitive probe of the crossover from a finite, strongly inhomogeneous droplet to a nearly homogeneous quantum fluid.

We have also shown that external confinement provides an independent way of controlling the acoustic response. Increasing the axial trapping strength compresses the droplet, enhances its central density, and increases the sound velocity. At sufficiently strong confinement, the collective properties continuously evolve toward those of a conventional trapped Bose gas. The excitation spectra obtained from linearization and the collective frequencies estimated using the sum-rule approach are consistent with this picture and provide independent characterizations of the low-energy response. The agreement between these approaches supports the interpretation of the real-time density-wave propagation as a genuine measurement of the acoustic response of the droplet.

Our results also suggest a feasible experimental route for observing sound propagation in low-dimensional quantum droplets. In particular, a two-component $^{39}$K Bose mixture provides an experimentally established platform for realizing self-bound droplets. A localized density depletion can be generated using a focused blue-detuned laser beam and released suddenly to launch two counter-propagating density waves. The characteristic propagation distances and times obtained in our simulations are within the range accessible to current ultracold-atom experiments, making a direct observation of the predicted acoustic dynamics realistic.

Several directions naturally follow from the present work. First, an experimental measurement of the propagation velocity would provide a direct test of the finite-size dependence predicted here and would establish a real-time probe of the equation of state of quantum droplets. Second, it would be interesting to investigate nonlinear sound propagation by increasing the amplitude or width of the initial density perturbation, where the linear sound-wave picture is expected to break down and shock-wave or soliton-like dynamics may emerge~\cite{engels2007stationary}. Third, finite-temperature effects could be incorporated to explore the modification of the acoustic response and the possible emergence of multiple sound modes~\cite{guebli2021quantum,spada2024quantum}. Finally, extending the present analysis to dipolar droplets, supersolid states, and polar molecular quantum droplets would allow one to investigate how anisotropic or long-range interactions modify sound propagation and could provide a broader connection between acoustic phenomena and emergent quantum-fluid phases.

%\textit{Notes added:}
%%%%%%%%%%%%%%%%%%%%%%%%%%%%%%%%%%%%%%%%%%%%%%%%%%
%%%%%%%%%%%%%%%%%%%%%%%%%%%%%%%%%%%%%%%%%%%%%%%%%%
\begin{acknowledgments}
We acknowledge useful discussions with Peng Zou. This work is supported by the Natural Science Foundation of China (Grants No. 12204413 and No. 12247101), the Science Foundation of Zhejiang Sci-Tech University (Grant No. 21062339-Y), the Fundamental Research Funds for the Central Universities (Grant No. lzujbky-2025-jdzx07), the Natural Science Foundation of Gansu Province (No. 25JRRA799), and the '111 Center' under Grant No. B20063.
\end{acknowledgments}

\emph{Data availability}---The data that support the findings of this article are openly available~\cite{yuan2026data}.

%%%%%%%%%%%%%%%%%%%%%%%%%%%%%%%%%%%%%%%%%%%%%%%%%%
%%%%%%%%%%%%%%%%%%%%%%%%%%%%%%%%%%%%%%%%%%%%%%%%%%
\appendix

\bibliography{sound1dQD}
\end{document}